\documentstyle[aps,epsf]{revtex}

\begin{document}
\twocolumn

\title{Comment on "New modes of halo excitations in the $^6$He
nucleus"}

\author{D.~V.~Fedorov, A.~Cobis, and A.~S.~Jensen}

\address{Institute of Physics and Astronomy, University of Aarhus, 8000
Aarhus C, Denmark}

\maketitle
\begin{abstract}
We try to explain the differences in the $^6$He dipole strength
function in \cite{dan97} and \cite{cob97}.  We perform the full basis
calculation of the strength function with the same renormalized
interaction as in \cite{dan97} and show that the size of the basis,
needed for converged calculations of the $^6$He continuum spectrum, is
much larger than that for the discrete spectrum.  The renormalized
interaction of \cite{dan97} therefore cannot be used for the continuum
spectrum calculations with the same basis as for the ground state.
\end{abstract}~\\

\narrowtext

In a recent paper \cite{dan97} the continuum properties of $^6$He were
discussed within a three body $nn\alpha$ model.  The strength functions
were computed using the hyper-spherical harmonics approach.  The
authors argued that the main features of the continuum spectrum of the
$nn\alpha$ system can be obtained using only a few hyper-spherical
harmonics with the hyper-spherical quantum number $K\leq 6$, if one
simultaneously renormalizes the $n\alpha$ interaction by introducing
additional attraction.

However the dipole strength function in \cite{dan97} differs
significantly from our recent calculation \cite{cob97}, where we used
the same three-body model.  The purpose of this comment is to explain
these differences.

We show that the continuum state calculations of the $nn\alpha$ system
demand a much larger basis, $K\leq 100$, compared to the bound state.
In such case the simple prescription to renormalize the $n\alpha$
interaction aiming at the bound state might not be adequate for the
continuum.

The {\bf method} we use is, in principle, similar to that of
\cite{dan97} -- the angular part of the wave function is expanded in
terms of the hyper-spherical harmonics and then the resulting system of
coupled hyper-radial equations is solved.  However our approach has the
following modifications \cite{fed94}:  i) We use Faddeev rather than
the Schr\"{o}dinger equation; ii) Prior to solving the hyper-radial
equations we calculate the eigenvalues $\lambda_n$, as functions of the
hyper-radius $\rho$, of the angular part of the Faddeev operator. The
quantities $\lambda_n(\rho)\rho^{-2}$ then serve as effective
potentials in the radial equations; iii) We use analytic large-distance
solutions of the Faddeev equations for short range potentials
\cite{jen97} which greatly enhances the method and allows fully
converged calculations.

The $n\alpha$ {\bf interaction}, originally fit to the scattering
data, was somewhat modified in \cite{dan97,dan91,dan93} in order to
compensate for the limited basis,
\begin{eqnarray}
&&{\hat V}_{n\alpha}(r)=\exp{(-r^2/2.35^2)}\times\\ 
&&(50\;{\hat P}_s-48.3\;{\hat P}_p-23.0\;{\hat P}_d
-11.71\;{\bf l\cdot s}).\nonumber
\end{eqnarray}
Here ${\hat P}_l$ is a projection operator on the corresponding orbital
momentum state l and {\bf s} is the neutron spin.
\begin{figure} 
\epsfxsize=3.3in \epsfbox{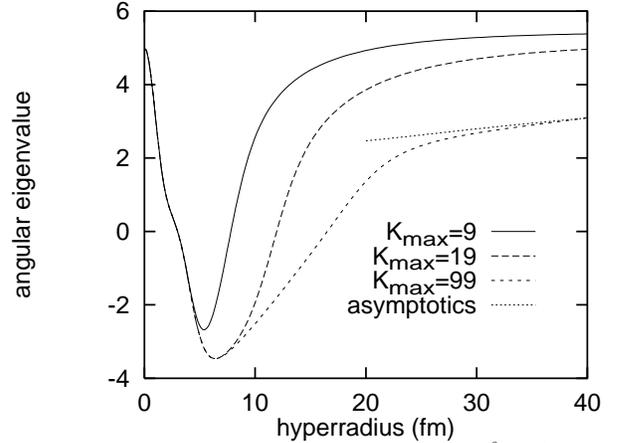}
\caption{ The lowest angular eigenvalue for $^6$He ($J^\pi=1^-$)
calculated with different $K_{max}$. Also shown is the eigenvalue
obtained from the asymptotic large distance expansion of the Faddeev
equations \protect\cite{jen97}.  }
\end{figure}
All lengths are in fm and energies in MeV.  Due to the over attraction
this interaction fails to reproduce the experimental $n\alpha$ phase
shifts.

For the $nn$ interaction we use a simple Gaussian \cite{gar96}, similar
to the the one used in \cite{dan91}. The potential reproduces the
experimental effective range in s-wave and scattering lengths in s- and
p-waves,
\begin{eqnarray}
&&{\hat V}_{nn}(r)=
\exp{(-r^2/1.8^2)}\times \\ 
&&(2.92+45.22\;{\bf s}_1{\bf s}_2-12.08\;{\bf l}\cdot({\bf s}_1
+{\bf s}_2)+26.85\;{\hat S}_{12}),\nonumber
\end{eqnarray}
where ${\bf s}_1$, ${\bf s}_2$ are the spins of the neutrons, ${\hat
S}_{12}$ is the usual tensor operator.

In Fig.~1 we illustrate the convergence of the lowest {\bf angular
eigenvalue} using, as examples, calculations with $K_{max}=9$,
$K_{max}=19$, and $K_{max}=99$.  Obviously the basis limited to $K<9$,
used in \cite{dan97}, is by far not converged.  It is only the curve
for $K_{max}\approx 100$ that finally converged towards the
large-distance asymptotics.  This long tail of the eigenvalue arises,
in the present case, due to large scattering length in the $nn$
subsystem \cite{jen97}. The scale of convergence towards the asymptotic
behavior is then around 20~fm.

The convergence problems for the lowest effective potential is
illustrated in Fig.~2.  The converged curve has the minimum value of
about 0.15 MeV and the low (0.3 MeV) barrier extending into the
asymptotic region beginning at about 25 fm.  In contrast the curve for
$K_{max}=9$ has minimum and maximum values at 0.7 MeV and 1.4 MeV,
respectively. A more realistic potential calculated with a correct
$n\alpha$ interaction \cite{cob97}, is accidentally close to the curve
for $K_{max}=19$.  

\begin{figure} 
\epsfxsize=3.3in \epsfbox{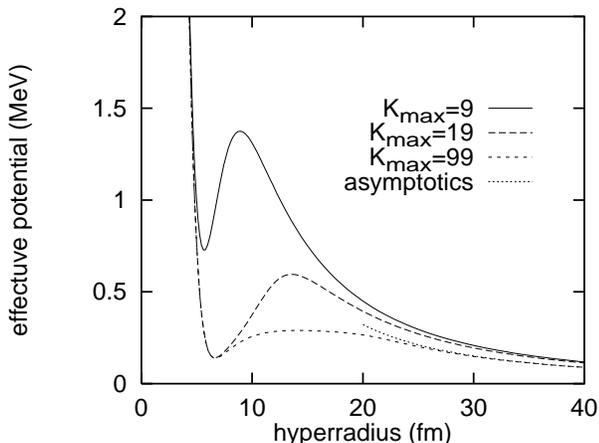}
\caption{ The lowest effective potential $(\lambda_1(\rho) +
15/4)\rho^{-2}\times$ $\hbar^2/(2m)$ for $^6$He ($J^\pi=1^-$)
calculated with different $K_{max}$ ($m$ is the nucleon mass). }
\end{figure}

Note that that the $K<9$ basis seems to be
sufficient for the $0^+$ ground state with the renormalized interaction
\cite{dan97}.  This shows that the renormalization technique needs
different bases for the ground state and the continuum state
calculations.

The {\bf dipole strength function} was calculated in \cite{dan97} with
a basis size corresponding to $K\leq 6$.  In Fig.~3 we show the results
of our calculations of the dipole strength function for different
$K_{max}$.  Our $K_{max}=9$ curve agrees quite well with \cite{dan97}.
Increasing $K_{max}$ causes the main peak to shift towards lower
energies with the second peak forming and also moving towards lower
energies.  The result for $K_{max}=9$ is rather smooth and deviates
relatively little from the corresponding result obtained with plane
waves for the 1$^-$ continuum wave function.  This leads to a
presumably misleading impression that there is little structure in the
$1^-$ continuum.  The other two curves, $K_{max}$=19 and $K_{max}$=99,
on the contrary show visible concentration of the strength at small
energies.

\begin{figure} 
\epsfxsize=3.3in \epsfbox{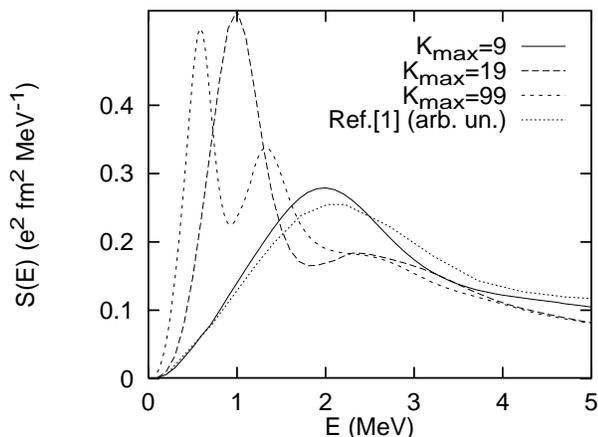}
\caption{ The dipole strength function $S(E)$ = $\sum_n\times$
$|\langle nE\| M(E1)\| gs\rangle|^2$ as a function of the three-body
energy $E$ for the transition from the ground state (0$^+$) to the
continuum state (1$^-$) of $^6$He calculated with different $K_{max}$.
The curve from \protect\cite{dan97} is given in arbitrary units.  }
\end{figure}

We have to note, however, that due to the additional attraction
introduced in the $n\alpha$ interaction in \cite{dan97} the converged
dipole strength function here is excessively shifted towards lower
energies and therefore is not realistic. The realistic dipole strength
function \cite{cob97} is close to the curve $K_{max}=19$.  In this case
the limited basis is approximately compensated by the over-attractive
potential.  Note that for the $0^+$ bound state calculations the
corresponding compensatory basis was $K_{max}=6$ \cite{dan97}.  This
shows that the continuum spectrum demands more careful and accurate
calculations.

In {\bf conclusion}, in order to explain the differences between the
dipole strength functions from \cite{dan97} and \cite{cob97} we have
performed a fully converged calculation of the $J^\pi=1^-$ continuum
spectrum of $^6$He using the same renormalized $n\alpha$ interaction as
in \cite{dan97}. We have shown that the number of basis functions
needed for converged calculations of the $^6$He continuum spectrum is
much larger than that for the discrete spectrum.  We conclude that the
technique of \cite{dan97} to renormalize the $n\alpha$ interaction to
compensate for the limited basis becomes ambiguous when applied to the
continuum spectrum.  The reason for the differences is therefore that
this renormalized interaction is not appropriate for continuum
calculations with the same basis size as for the ground state.

\end{document}